# Decentralized Certificate Authorities

Bargav Jayaraman, Hannah Li, David Evans*
University of Virginia
https://oblivc.org/dca
[bargavj,hannahli,evans]@virginia.edu

## ABSTRACT

The security of TLS depends on trust in certificate authorities, and that trust stems from their ability to protect and control the use of a private signing key. The signing key is the key asset of a certificate authority (CA), and its value is based on trust in the corresponding public key which is primarily distributed by browser vendors. Compromise of a CA private key represents a single point-of-failure that could have disastrous consequences, so CAs go to great lengths to attempt to protect and control the use of their private keys. Nevertheless, keys are sometimes compromised and may be misused accidentally or intentionally by insiders. We propose splitting a CA's private key among multiple parties, and producing signatures using a generic secure multi-party computation protocol that never exposes the actual signing key. This could be used by a single CA to reduce the risk that its signing key would be compromised or misused. It could also enable new models for certificate generation, where multiple CAs would need to agree and cooperate before a new certificate can be generated, or even where certificate generation would require cooperation between a CA and the certificate recipient (subject). Although more efficient solutions are possible with custom protocols, we demonstrate the feasibility of implementing a decentralized CA using a generic two-party secure computation protocol with an evaluation of a prototype implementation that uses secure two-party computation to generate certificates signed using ECDSA on curve secp192k1.

## 1 INTRODUCTION

The security of the web heavily relies on anchoring trust on the certificate authorities. Although widely-trusted CAs take extraordinary efforts to protect their private keys, compromises and abuses still occur, with disastrous consequences. Roosa and Schultze [56] discussed attacks on certificate authorities where the attacker either takes control of the CA or colludes to obtain certificates trusted by a client's browser. As an example, an attack on Comodo's registration authority (RA) allowed the attacker to obtain a domain validation (DV) certificate for Google [58]. An attacker exploited a subCA of DigiNotar to obtain several extended validation (EV) certificates [57]. EV certificates provide access to high security HTTPS websites, and hence their leakage is more serious compared to the leakage of DV certificates.

Certificate authorities are at risk of not just compromise, but also internal misuse and mistakes. Notably, Symantec has issued certificates without carefully verifying the domain and expiration date or reviewing the certificate recipient's application [48].

In 2015, Symantec mistakenly issued test certificates for a number of domains, including www.google.com, without the permission of the domain holders. In response to suspicion that Symantec has allegedly mis-issued more than 30,000 certificates, Google Chrome recently announced plans on restricting certificates issued by Symantec-owned issuers [27]. Free open certificate authority Let's Encrypt, which provides free certificates to requesters using an automated domain validation protocol, has been criticized for issuing TLS/SSL certificates to 15,270 sites containing the keyword paypal.com which were obtained for use in phishing attacks [60].

The above attacks are possible due to single point-of-failure where the certificate signature generation depends on a single private key owned by a single CA. This key is stored in a physical file that could be compromised by an outside attacker exploiting software vulnerabilities, by an attacker who obtains physical access to the stored key, or by insiders who have access to it within the organization. Our goal is to provide a mechanism to reduce the single point-of-failure risks inherent in current CAs. Our solution is to split the private signing key into shares. The private key is never reconstructed in any plaintext form. Instead, a new certificate is generated using a multi-party computation protocol where the private key is only reconstructed in encrypted form within a secure protocol. Secure multi-party computation (MPC) enables two or more parties to collaboratively compute a function without revealing their individual inputs. We use MPC to enable a key signing operation that depends on a private key, without ever exposing that private key. The key itself is never visible to any party in unencrypted form, and never stored on any physical device.

The shares could be distributed among multiple hosts operated by the same CA to mitigate the risk of compromise when one of the keys is stolen. To further reduce mistakes and insider risks, the shares could be distributed across two independently-operated CAs with their own procedures of vetting so that scrutiny from multiple parties would reduce the chance of mistakes such as those made by Symantec [48] and Let's Encrypt [60]. The final scenario supports organizations that do not want to place their identity in the hands of a CA. The private key is split between the CA and the organization, requiring involvement of both parties to generate the joint public key the first time and a new certificate at each subsequent signing. This joint public key is signed by a pair of two independent CAs. This design allows organizations to control the generation of any certificate for their domain, and prevents a CA from issuing certificates to third-parties without the permission of the domain holders. However, adopting our scheme would require some modifications to the current certificate deployment model. For first two scenarios, where the organization is not involved in its certificate signing, the joint public key used in the signing can be baked into the client browser to verify the certificate signature. For the third scenario, where the organization is also involved in the signing, baking the

---

joint key into the client browser can be expensive, as there could be many such keys, one for each organization. We could employ certificate transparency logs [39] and public key pinning [14] to solve this problem. Further discussion regarding this can be found in Section 6.

In this work, we demonstrate the practicality of a decentralized CA by implementing certificate signing as a secure two-party computation using the Elliptic Curve Digital Signature Algorithm (ECDSA) on a 192-bit curve. Although there are more efficient ways to perform multi-party ECDSA signing as a specific protocol (see Section 7), we believe it is useful to see how feasible it is do to this using a generic secure computation protocol. By using a generic protocol, it is easy to adapt the signing to other protocols or to incorporate other computations such as policy checks in the signing process, as well as to support a variety of privacy models. The main cost of signing depends on the cost of bandwidth between the participating CAs. If both are in the same data center where in-center bandwidth is free, the computational cost to sign one certificate on Amazon AWS EC2 c4.2xlarge nodes using Yao's garbled circuits protocol with semi-honest security is 28.2-32.6 cents; for active security using dual execution, this cost approximately doubles to 63.3-65 cents. Performing the signing with dual execution involves transferring 819.44 GiB, so at AWS prices for bandwidth this makes the cost per certificate up to $17.07 for active security when nodes are running in different data centers. Although this cost may be prohibitive for some purposes, for others (e.g., Symantec currently charges a $500/year premium for ECC over RSA/DSA certificate) it is already economically viable.

**Roadmap.** We begin by providing a background on the certificate issuance process (Section 2.1), secure multi-party computation (Section 2.3), and the ECDSA signing algorithm (Section 2.2). In section 3, we explain the design of our decentralized certificate authorities. Section 4 provides details on our implementation, followed by the performance measurements in Section 5. Finally, we end our paper discussing the adaptability of our design in Section 6.

## 2 BACKGROUND

This section provides background on how certificates are issued, secure multi-party computation protocols, and the ECDSA signing algorithm.

### 2.1 Certificate Issuance Process

Digital certificates are issued by trusted certificate authorities for certificate recipients (more formally referred to as *subjects*) who wish to establish a secure connection with the clients visiting their web page. Here, we provide some background on the contents and format of X.509 digital certificates used in TLS, and the process followed to issue a certificate.

**X.509 Digital Certificates.** The standard format of TLS/SSL public key certificates is X.509, which is made up of three components: (1) the certificate, which contains information identifying the issuer and the subject and the scope of the certificate (version number and validation period), (2) the certificate signature algorithm specifying the algorithm that was used for signing, and (3) the certificate signature [16]. Whereas verifying the parameters of the certificate component (1) is important in our decentralized CAs set up, our primary goal is to compute the certificate signature (3) using a joint signing process where the signing key is never exposed.

**Subject Requests for a Certificate.** In the current X.509 setup, the subject must apply for a certificate through submitting a *Certificate Signing Request* (CSR) [64]. The first time the subject requests a certificate, it generates a public-private key pair, including the public key in the CSR and using the private key to sign the CSR. Note that this key pair and signature is different from the CA's key pair and certificate signature that must be generated using the CA's private key. The subject's signature provides a consistency check that allows the CA to verify the CSR signature with the subject's provided public key.

**Verifying the Subject's Identity.** Today, attesting to the applicant's CSR contents has become increasingly automated through ad hoc mechanisms, though manual efforts still remain. Through the open-source automatic certificate management environment (ACME) (as used by Let's Encrypt), the applicant can prove its ownership of the domain by responding to a series of server-generated challenges (i.e., proving control over a domain by setting a specific location on its website to an ACME-specified string) [28]. ACME allows certificates to be issued at a lower cost, but the lack of manual scrutiny allows phishing websites, for instance, to obtain certificates for popular domain look-alikes, which happened with paypal.com [60]. It also only attests to having (apparent) control of the domain at the time of the request, but does not provide any way to verify that it is the legitimate owner of that domain or a trustworthy organization.

The attestation process for proprietary CAs is not as publicly known as ACME, and often includes non-automated steps. One of the manual validation steps completed by Comodo is a phone call to a number verified by a government or third party database, or by legal opinion or accountant letter. This extra manual effort both would have prevented the paypal.com incident and stopped an attacker with a stolen CSR private key and access to the victim's server from obtaining a certificate.

Certificates designated as *Extended Validation* (EV) involve more extensive validation. In order to issue EV certificates, a CA must pass an independent audit [13] and follow defined criteria in issuing certificates including verifying the subject's identity through a stricter, manual vetting process that includes both the Baseline Requirements and the Extended Validation requirements [66]. Major browsers display the padlock and green address bar for EV certificates. This conveys a higher level of security to customers, and CAs charge subjects a premium to obtain them.

**CA signs the Certificate.** After verifying the subject's identify and validity of the request, the issuer composes the (1) certificate. Although we colloquially refer to the issuer's role as simply 'signing a certificate', the technical term for the input to the signature algorithm is the *TBSCertificate*, which is actually the (1) certificate. The TBSCertificate is very similar to the CSR. But, whereas the CSR only contains information about the subject, the TBSCertificate contains pertinent information about both the subject and issuer. The output to the signature function is the (3) certificate signature. In the rest of the paper, we frequently use the terms *signature generation* and *certificate generation* interchangeably to mean the process whereby a signed certificate is created from the TBSCertificate contents.

## 2.2 Elliptic Curve Digital Signature Algorithm

ECDSA is a popular certificate signing algorithm that is included with most TLS libraries including OpenSSL and GnuTLS, and is supported by major web browsers including Mozilla Firefox, Google Chrome, Internet Explorer, Safari and Opera.

ECDSA depends on modular arithmetic operations on elliptic curves defined by the equation

$$y^2 \equiv x^3 + ax + b \mod p$$

The curve is defined by three parameters: $p$, a large prime defining the curve finite field $\mathbb{F}_p$, and the coefficients, $a$ and $b$. The size of the parameters defines the size of the elliptic curve. NIST specifies curves recommended for ECDSA with 192 bits, 256 bits, 384 bits and 512 bits [53].

Signing depends on three additional public parameters as defined by Standards for Efficient Cryptography Group (SECG) [15]: $G$ is the base point of the curve, $n$ is the order of $G$, and $h$ is the cofactor. All the parameters are of same size as $p$.

Signing requires a private-public key pair $(sk, pk)$. The private key $sk$ is chosen uniformly randomly from the range $[1, p-1]$, and the corresponding public key $pk$ is obtained from the curve point multiplication $pk = sk \times G$.

Signing begins with generation of a uniform random number, $k$, in the range $[1, p-1]$, which is used to obtain a curve point $(u, v) = k \times G$. One part of signature is $r = u \mod n$. At this stage, if $r$ is 0 (which should happen with negligible probability), then the whole process is repeated with freshly generated random number $k$. The other part of signature is $s = k^{-1}(z + rd) \mod n$, where $d$ is the private key used for signing and $z$ is the hash of the message to be signed. Again, a check is performed to verify that $s$ is not 0. If it is, the process is repeated with freshly generated random number $k$. Otherwise, the final pair $(r, s)$ is the signature, which is made public.

## 2.3 Secure Multi-Party Computation

Secure multi-party computation enables multiple parties to collaboratively compute a function on their private inputs without revealing any information about those inputs or intermediate values. At the end of the protocol, the output of the computation is revealed (to one or more of the parties). We limit our focus here to two-party computation because it fits our problem setting well, although several methods exist for extending multi-party computation to more than two parties. Two-party computation was introduced by Yao [68], and he presented a generic method for computing any function securely using garbled circuits in a series of talks in the 1980s that is now known as "Yao's protocol". The protocol enables any function that can be represented by a Boolean circuit to be computed securely.

In a standard Yao's protocol execution, one party, known as the *generator*, generates the garbled circuit that computes over encrypted inputs and passes it to the other party, known as the *evaluator*, who evaluates the circuit by plugging in the encrypted inputs. Instead of operating on semantic values, the garbled circuit operates on wire labels, random nonces that provide no information, but allow the evaluator to evaluate the circuit. The evaluator obtains the wire labels corresponding to her own inputs using an oblivious transfer protocol [22], and extended OT protocols enable large inputs to be transferred efficiently [29, 33, 49, 51]. The output of the computation is revealed to the parties at the end of the protocol, typically by providing hashes of the final output wire labels to the evaluator, who sends the semantic output back to the generator.

Yao's protocol provides security only against *semi-honest* adversaries, which are required to follow the protocol as specified but may attempt to infer private information from the protocol transcript. It does not provide security against active adversaries since there is no way for the evaluator to know she is evaluating the correct circuit, and there is no way for the generator to verify the output provided by the evaluator is the actual output. A malicious generator could generate a circuit that computes some other, more revealing, function on the other party's input, instead of the agreed-to function. In our scenario, instead of outputting the correct signature (which should not reveal the other party's key share), the circuit could be designed to output the other party's key share masked with a value known to the generator.

Hence, although semi-honest security can provide some protection by avoiding the need to store the key in a way that could be exposed, is not sufficient for implementing a decentralized CA in any adversarial context. The most common methods for hardening MPC protocols to provide security against active adversaries is to use like cut-and-choose [42]. In a standard cut-and-choose protocol, the generator generates $k$ garbled circuits and the evaluator chooses a fraction of the circuits and asks the generator to reveal those circuits to test their correctness. The evaluator then uses the remaining circuits for secure computation. Although there have been many improvements to the basic cut-and-choose protocol [1, 2, 11, 25, 31, 36, 40, 43–46, 52, 55, 62, 63, 67], it remains expensive for large circuits.

**Dual execution.** The alternative, which we explore in this work, is to relax the strict requirements of fully malicious security to enable more efficient solutions. In particular, we use dual execution [30, 47], which provides active security against arbitrary adversaries, but sacrifices up to one bit of leakage to provide a much more efficient solution. In a dual execution protocol, Yao's protocol is executed twice, with the parties switching roles in the consecutive executions. Before revealing the output, though, the outputs of the two executions are tested for equality using a malicious-secure equality test protocol. The output is revealed to both the parties only if the equality check passes. Otherwise, it is apparent that one of the parties deviated from the protocol and the protocol is terminated. A malicious adversary can exploit this to leak a single bit that is (at worst) the output of an arbitrary predicate by using selective failure—designing a circuit that behaves correctly on some inputs but incorrectly on others, so the equality test check result reveals which the subset of inputs contains the other party's input.

We use Yao's protocol in the case of semi-honest adversaries and dual execution in the case of malicious adversaries. In the next section, we discuss our signing algorithm and how we adopt these protocols for our scenario.

## 3 DESIGN

Implementing a decentralized CA requires protocols for generating a joint signing key (without ever disclosing the private key to any party) and for signing a certificate using a split key. Both of these protocols use secure multi-party computation to allow the two hosts[1] that are combining to implement a joint CA to securely compute these functions while keeping their private inputs secret.

We assume that the two hosts have already agreed on the signing algorithm and public elliptic curve parameters (ECDSA in our prototype implementation), and have decided on the methods they will use to validate certificate requests. These methods can be independent, indeed it is best if they are. All that is necessary is that both participants must agree before any certificate can be signed using the joint signing key.

Next, we present the key generation and signing protocols. Section 3.3 provides a brief security argument. Section 3.4 discusses different ways our protocols could be used.

### 3.1 Key Pair Generation

Before a pair of hosts can perform joint signings, they need to generate a joint public-private key pair. This is a one-time process and is independent of the certificate signing protocol. At the end of this process, a joint public key is published and each host has a share of the corresponding private key. The private key never exists in any unencrypted form, so the only exposure risks are if *both* hosts are compromised or the MPC protocols are corrupted.

Figure 1 illustrates the key pair generation protocol. To begin, each host generates a random value that will be its private key share: $sk_A$ (owned by party $A$) and $sk_B$ (owned by party $B$). These are drawn uniformly randomly from the range $[0, p-2]$. The master private key, $sk = (sk_A + sk_B) \mod (p-1) + 1$, is in the range $[1, p-1]$. Because of the properties of modular addition, this key will be uniformly randomly distributed in $[1, p-1]$, even if one of the input keys is not. The private key, $sk$, is not revealed to either of the parties and will only be represented by encrypted wire labels inside the MPC signing protocol.

Within the MPC, the key shares are combined to produce $sk$ as described above, and the elliptic curve point multiplication is performed to generate the public key $pk = sk \times G$. This value is revealed as the output to both the parties. Section 4.2 provides details on how curve point multiplication is implemented, which is used in both key generation and signing.

### 3.2 Signing a Certificate

Once a pair of hosts have generated a joint key pair, they can jointly sign certificates using the shared private key without ever exposing that key. The inputs to the signing protocol are the key shares ($sk_A$, $sk_B$), the random number shares ($k_A$, $k_B$), and the message to be signed, $z$. The key shares are those that were used as the inputs to the key pair generation protocol to produce the joint key pair.

Each signing server generates an $L_n$-length random integer in the range $[0, p-2]$ that will be its share of the random number $k$, needed for ECDSA signing. These shares, $k_A$ and $k_B$, are inputs to the MPC protocol and will be combined to obtain $k = (k_A + k_B) \mod (p-1) + 1$ (which is uniform in the range $[1, p-1]$) inside the protocol so neither party can control $k$.

Both hosts must agree on the same TBSCertificate to be signed, and use whatever out-of-band mechanisms they use to validate that it is a legitimate certificate request. As specified by ECDSA, the input to the signing function, $z$, is the $L_n$ left-most bits of *Hash*(*TBSCertificate*), where $L_n$ is bit length of the group order $n$ [65]. Both parties compute $z$ independently, and feed their input into the MPC (these are denoted as $z_A$ and $z_B$ in Figure 2). There is no secret involved in computing $z$, but it is essential that both parties know and agree to the certificate they are actually signing. Hence, both parties provide their version of $z$ as inputs to the secure computation, so there is a way to securely verify they are identical.

The signing algorithm first compares the inputs $z_A$ and $z_B$ to verify that the two parties are signing the same certificate. Then, $k$ and $sk$ are obtained by combining the shares from both the parties as explained above. The remaining steps follow the ECDSA signing algorithm mentioned in Section 2.2. At the end of the protocol, the certificate signature pair $(r, s)$ is revealed to both parties. In rare instances, the signing algorithm may result in $r = 0$ or $s = 0$, in which case the whole process is repeated by both the parties with different values of $k_A$ and $k_B$. To avoid any additional disclosure in the dual execution model, however, it is necessary to complete the signing process even when $r = 0$ and perform the secure equality test on the result so there is no additional leakage opportunity for an adversarial generator.

### 3.3 Security Argument

The above signing algorithm is implemented using both Yao's protocol and dual execution. The security of our protocols follows directly from the established security properties provided by the underlying MPC protocols we use, so no new formal security proofs are needed.

For semi-honest Yao's protocol, which we do not recommend but include in our cost analysis for comparison purposes, one party acts as the circuit generator and the other party acts as the circuit evaluator. The generator creates a garbled circuit encoding the above signing algorithm and passes it to the evaluator along with the wire labels corresponding to his inputs. The evaluator obtains the wire labels for its inputs using extended oblivious transfer, and gets the result of the computation. For instance, if party $A$ is the generator, then $A$ passes the circuit along with the wire labels corresponding to $k_A$, $sk_A$ and $z_A$ to party $B$, who acts as the evaluator. Party $B$ obliviously requests the wire labels corresponding to $k_B$, $sk_B$ and $z_B$ from $A$, and evaluates the circuit, which first asserts $z_A == z_B$. Revealing the assertion failure at this point leaks no information, since it only depends on the non-secret $z$ inputs. If the assertion passes, it proceeds to obliviously combine the input shares to compute the $k$ and $sk$ to be used in the signing. A sanity check is also embedded to ensure that the shares $k_A$, $sk_A$, $k_B$ and $sk_B$ are in the proper range (although a host that provides invalid inputs would not obtain any advantage since they are combined with the other host's random shares using modular arithmetic). The circuit computation finally outputs the signature pair $(r, s)$, or outputs an error message if either $r = 0$ or $s = 0$. The output is revealed to both parties.

---
[1] We use *hosts* in this presentation as a generic term for the participant in a protocol. As we discuss in Section 3.4, in our case, the hosts could be separate servers operated by a single CA, two separate servers operated by independent CAs, or independent servers operated by a CA and a subject.

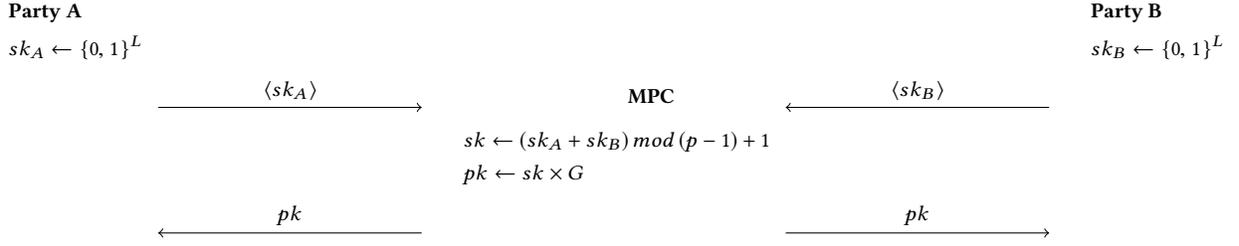

**Figure 1: Protocol for generating key pair,** GENERATEKEYPAIR: $sk_A; sk_B \to pk; pk$. Party A generates one share of secret key, $sk_A$, and Party B generates other share, $sk_B$. The output of protocol is public key, $pk$, that pairs with $sk = sk_A \oplus sk_B$.

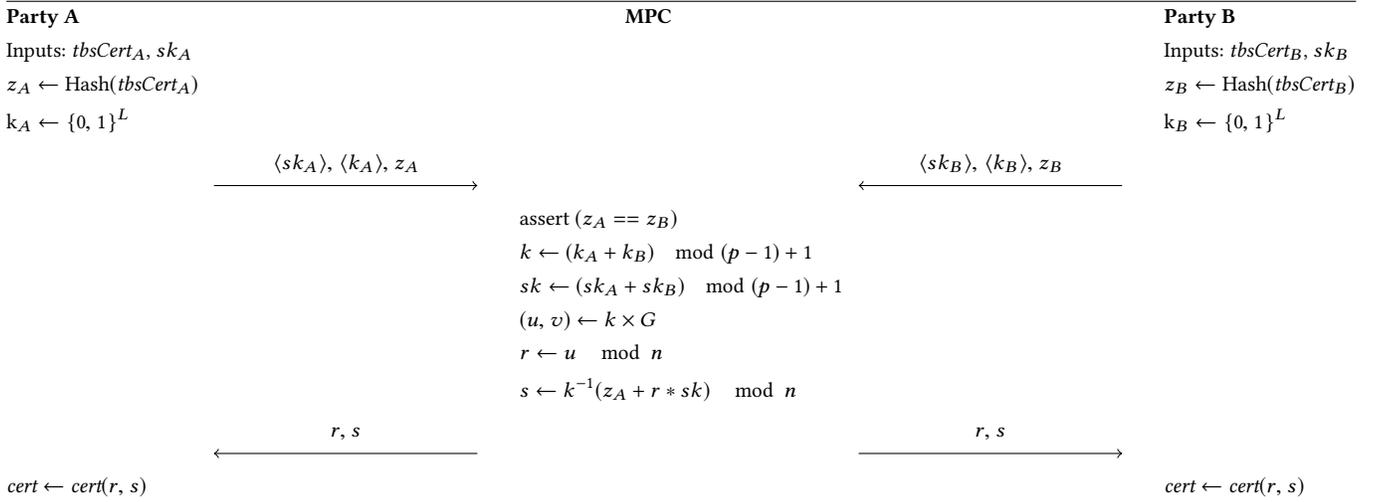

**Figure 2: Protocol for generating certificate signature,** GENERATECERTIFICATESIGNATURE: $(sk_A, k_A, z_A); (sk_B, k_B, z_B) \to (r, s); (r, s)$. Party A generates one share of the random integers, $k_A$, and Party B generates other random integer, $k_B$. The output of protocol is certificate signature, $(r, s)$, signed with secret key shares $sk_A$ and $sk_B$. At the start of this signing protocol, we assume that the two parties have already agreed to issue the certificate and on the contents of it.

Since, the semi-honest protocol is not secure against malicious adversaries, a malicious generator can manipulate the circuit to produce a bad certificate (i.e., obtain signature for a different certificate than for the agreed-upon message $z$) or to leak bits of the other party's secret key shares in the output (which could still appear to be a valid signature on $z$). An easy attack would be to generate a few different valid signatures with low fixed values of $k$ (which could be done with small sub-circuits), and then select among these in a way that leaks bits of the other host's private key share.

For these reasons, we do not think it is advisable to perform something as sensitive as certificate signing using only semi-honest security. Instead, we use the dual execution protocol to provide active security with a single bit leakage risk. With dual execution, Yao's protocol is executed twice with both the parties switching their roles in the consecutive runs, i.e. if party $A$ is the generator in the first execution, it then becomes the evaluator in the next execution. However, the main difference is that the output of both the runs is revealed only after passing a maliciously-secure equality test that ensures that the semantic outputs of both the executions are identical. This ensures that neither of the parties deviated from the protocols and, as a consequence, generated a bad certificate. We use a maliciously secure extended OT protocol [33] to obtain the inputs, so the only leakage possibility is the one-bit selective failure predicate.

In some applications, the risk of leaking an arbitrary one-bit predicate on private inputs would be severe. For our key signing protocol, however, it is negligible. The best a malicious adversary can do is leak a single bit of the other party's private key share for each execution, with a probability of $\frac{1}{2}$ of getting caught. The signing algorithms used have sufficient security margins that leaking a few bits is unlikely to weaken them substantially, and the participants in a joint CA have a strong incentive to not be caught cheating.

### 3.4 Scenarios

We consider using our decentralized CA in three different settings: jointly computing the signature and public key by (1) a single CA that wants to protect its own signing keys by distributing them among its own hosts, (2) two independent CAs who want to jointly mitigate the risks of key exposure or misuse, and (3) a CA and subject who wants to protect itself from a CA generating rogue certificates by being directly involved in the certificate signing process.

The first two scenarios are identical technically, just differ in the out-of-band processes that are used to validate a certificate request. In the first scenario, the risk of key exposure is reduced by never storing the key in a physical form in one place. The two hosts with key shares can be placed in different locations, protected by different physical mechanisms and people. The decision to generate a certificate is still made in a centralized way, and a single organization controls both key shares.

In the second scenario, the two hosts are controlled by different organizations. Each could implement its own policy and methods for validating a certificate request. This reduces the risks of mistakes and insider attacks since people in both organizations would need to collude (or be tricked) in order for such an attack to succeed. By agreeing to a joint signing process, the two organizations could also enter an agreement to share liability for any inappropriate certificates signed by the joint key.

In the third scenario, the subject, $S$, holds the power to creating its own certificate by owning a share of the private key that is used to generate the signature. The subject and $CA_A$ first jointly create a public key using the key generation protocol. In order of this newly generated public key to be trusted, a certificate is needed that vouches for this public key as part of a key pair which is jointly owned by $CA_A$ and $S$. To support decentralization, this certificate should be signed by a joint CA comprising two different CAs, $CA_B$ and $CA_C$. Both $S$ and $CA_A$ will make independent requests to $CA_B$ and $CA_C$ to sign a joint certificate that associates the generated $pk$ with subject $(S, CA_A)$. After validating these requests, $CA_B$ and $CA_C$ will perform a joint signing protocol to generate a signed certificate that attests to $pk$ being the public key associated with the joint $(S, CA_A)$ entity. With this, $S$ and $CA_A$ can jointly produce a new certificate that attest to a new key pair, which is validated by the certificate chain from the joint $(CA_B, CA_C)$ certificate authority. For this scenario to add value, it is important that browsers trust the joint root key of $(CA_B, CA_C)$, but not fully trust the root key of $CA_A$ (otherwise, $CA_A$ could still generate its own certificates attesting to $S$). (We discuss deployment issues further in Section 6.)

## 4 IMPLEMENTATION

For our multi-party computation, we utilize the Obliv-C framework [70] that includes the latest enhancements to Yao's garbled circuits protocol [69] including garbled row reduction [50, 55], free XOR [35], fixed-key AES garbling [8] and half-gates [71]. These reduce the number of ciphertexts transmitted per non-XOR gates to 2, while preserving XOR gates with no ciphertexts or encryption operations. Apart from the Yao's protocol [68, 69] for garbled circuits that is secure against semi-honest adversaries, Obliv-C also implements the dual execution protocol [30] to provide security against malicious adversaries. The Obliv-C framework compiles an extended version of C into standard C, converting an arbitrary computation into a secure computation protocol by encoding it into a garbled circuit. It achieves performance executing two-party garbled circuits comparable to the best known results, executing circuits approximately 5 million gates per second over a 1Gbps link.

### 4.1 Big Integer Operations

Since Obliv-C is based on C, it naturally extends all the basic C data types and the operations over them. However, ECDSA requires computation over large curve parameters of size in the range of 192, 256, 384 and 512 bits. We use the Absentminded Crypto Kit [20], which is built over Obliv-C framework, to support secure computation over arbitrarily large numbers. We implement ECDSA over the standard NIST curve secp192k1 [15], although our implementation can be easily adopted for any standard elliptic curve. While the Absentminded Crypto Kit provides basic operations like addition, subtraction, multiplication and division over arbitrarily large numbers, the ECDSA algorithm requires modular arithmetic operations. Hence we are required to handle the underflows and overflows in the intermediate computations, thereby increasing the number of invocations of basic arithmetic operations over big numbers. This impacts the overall computation cost (we discuss the cost of these individual operations in Section 5, Table 2).

### 4.2 Curve Point Multiplication

The most expensive step in ECDSA (for both key generation and signing) is the curve point multiplication protocol. In order to calculate $k \times G$ efficiently, where $k$ is a positive integer and $G$ is the elliptic curve base point, we implement the double-and-add procedure (Algorithm 1), which invokes the pointAdd and pointDouble subroutines $L_n$ number of times at the most, where $L_n$ is the number of bits of $k$. In comparison, a naïve curve multiplication procedure would require $k$ invocations of pointAdd, where $k$ is a large value in the range $[1, p-1]$, where $p$ is the curve parameter of the order $2^{L_n}$.

Normally, the double-and-add procedure is prone to timing side-channel attacks since the number of executions of the if block reveals the number of bits set in $k$. Within an oblivious computation, however, the actual value of the predicate is not known and all paths must be executed obliviously. This means there can be no timing side-channels since all of the executing code is operating on encrypted values. In our implementation, this is made explicit with the **obliv if** in Obliv-C, which must be used when the predicate involves oblivious values. It obliviously executes both branches, but ensures (obliviously) that the computation inside the block takes effect only if the predicate was true.

The procedures pointAdd and pointDouble output the resulting curve point $R$, following the equations in Figure 3. All the operations are modulo $p$, where $p$ and $a$ are the elliptic curve parameters explained earlier. Here $(N_x - Q_x)^{-1}$ and $(2N_y)^{-1}$ are modular inverse

**Algorithm 1:** Double-and-Add Based Point Multiplication

1 <u>curveMult</u> $(k, G)$;
2 $N \leftarrow G$;
3 $Q \leftarrow 0$;
4 **for** *bit position* $i \leftarrow 0$ *to* $L_n$ **do**
5     **if** $k_i = 1$ **then**
6         $Q \leftarrow \text{pointAdd}(Q, N)$;
7     **end**
8     $N \leftarrow \text{pointDouble}(N)$;
9 **end**
10 **return** $Q$;

| $\text{pointAdd}(Q, N) \rightarrow (R_x, R_y)$ | $\text{pointDouble}(N) \rightarrow (R_x, R_y)$ |
|---|---|
| $\lambda = (N_y - Q_y) \times (N_x - Q_x)^{-1}$ | $\lambda = (3N_x^2 + a) \times (2N_y)^{-1}$ |
| $R_x = \lambda^2 - Q_x - N_x$ | $R_x = \lambda^2 - 2N_x$ |
| $R_y = \lambda(Q_x - R_x) - Q_y$ | $R_y = \lambda(N_x - R_x) - N_y$ |

**Figure 3: Eliptive Curve Point Operations**

operations, implemented efficiently using the Extended Euclidean algorithm.

### 4.3 Dual Execution

In dual execution, the protocol must follow feed-compute-reveal flow, such that all the inputs are required to be fed into the garbled circuit in the beginning and only then the computation can be performed, and finally the output is revealed in the end. This is important to satisfy the requirements of the dual execution security proof. Thus, it is not valid to feed in some input during the computation or to reveal any output before the completion of computation. Hence, in the ECDSA algorithm, we cannot reveal to the parties if $r = 0$ or $s = 0$ while the computation is in progress. An exception to feed-compute-reveal flow is the assertion of TBSCertificate hashes generated by both the parties $z_A == z_B$ in the beginning, where both $z_A$ and $z_B$ are in plain text and hence are not part of the garbled circuit computation. This allows us to break the computation in case the assertion fails, and does not breach privacy since the message hashes are public values.

## 5 COST

The goal of the cost evaluation is to understand the actual cost of deploying a decentralized CA in different scenarios. Since latency of key generation does not matter much, and it is easy to parallelize MPC executions to reduce latency [12, 32], we focus our analysis on the financial cost of operating a decentralized CA. Except in cases where inter-host bandwidth is free, this cost is dominated by bandwidth. We report results for signing only, since the key generation task is less expensive than signing, and is only done once for each key pair, where we expect many signings to be done.

**Setup.** Our experiments were performed on Amazon Elastic Compute Cloud (EC2) using c4.2xlarge nodes with 15 GiB of memory and 4 physical cores capable of running 8 simultaneous threads. We selected c4.2xlarge nodes as they are the latest compute-optimized nodes of AWS and have a dedicated EBS bandwidth of 1000 Mbps. For the operating system, we used Amazon's distribution of Ubuntu 14.04 (64 bit).

We implement the NIST elliptic curve secp192k1 [15], which has 192-bit parameters. All the private and random keys shares are 192-bit numbers generated using C's GNU MP Bignum Library [23] (GMPLib). We use Obliv-C [70] for secure MPC and our garbled circuit implementation uses 80-bit long cryptographic random wire labels. Secure computation over the 192-bit parameters is done using Absentminded Crypto Kit [20]. The hashing algorithm used for our input to the signing algorithm is SHA256, and our TBSCertificate is generated using OpenSSL [24] using a sample certificate data that Symantec could have used to sign Google's certificate. Regardless of the data contained in the TBSCertificate, the SHA256 hashed TBSCertificate results in a 32-byte string that is later submitted to the signing function. The TBSCertificate appears in the final certificate and is used by browsers to verify the signature. All of the data are fed into the MPC as 8-bit integer arrays.

We evaluated the signing costs with three different scenarios of servers/nodes based on the distance between them: *Local*, where both the parties run MPC on same machine; *Same Region*, where each party runs on separate machines in the same region (we used US East – Northern Virginia) and *Long Distance*, where each party runs on separate machines in different data centers; we used one host in US East – Northern Virginia and the other in US West – Northern California. This experimental setup helps us know to what extent the decentralized certificate authority is practical when the participating CAs are geographically separated.

**Results.** Table 1 summarizes the runtime measurements and cost estimates for signings completed using both the semi-honest and dual execution protocols for the three scenarios.

To simulate running a real server, the nodes were loaded with increasing number of simultaneous signings until the lowest compute time per signing was discovered. The compute time per signing is the largest when the two servers are far from each other. We measured the bandwidth between two EC2 nodes using iperf and found it to be 2.57 Gbits/sec between two hosts in the same region (US East - Northern Virginia) and 49.8 Mbits/sec between hosts in US East – Northern Virginia and US West – Northern California data centers.

The optimal setting for Local signing with Yao's protocol was for 24 parallel signings, which took 17 hours to produce 24 signatures, costing 28.2 cents per signing at the current AWS on-demand price for a c4.2xlarge node of 39.8 cents per hour. Dual execution achieved the best results at 16 parallel signings taking 25.5 hours to do so and costing 63.3 cents per signing. For nodes in the same region, Yao's protocol performed the best for 32 parallel signings, taking 14.1 hours and costing 32.6 cents per signing. Whereas, dual execution performed the best for 24 parallel signings, taking 19.6 hours and costing 65 cents per signing. These results are as expected, since the extra latency between the nodes means additional parallel processes can take advantage of delays waiting for network traffic.

The cost of bandwidth within an AWS data center is free, so for the Local and Same Region scenarios, only computation cost matters. For cross-country signings, we found best latency results with 72 parallel signings for Yao's protocol and 40 parallel signings for dual execution. The cost for this scenario is dominated by the cost of bandwidth, since circuit execution requires transmitting over 400 GiB of data. At AWS bulk transfer cost of 2 cents per GB across different regions of AWS, the total cost per signing came as $8.54 for Yao's protocol and $17.07 for dual execution.

|                                       | Local |        | Same Region |        | Long Distance |        |
|---------------------------------------|-------|--------|-------------|--------|---------------|--------|
|                                       | Yao   | DualEx | Yao         | DualEx | Yao           | DualEx |
| Optimal # simultaneous                | 24    | 16     | 32          | 24     | 72            | 40     |
| Compute Time (hrs)                    | 17.0  | 25.5   | 13.1        | 19.6   | 31.3          | 34.3   |
| Duration (/signing)                   | 0.708 | 1.59   | 0.409       | 0.817  | 0.435         | 0.858  |
| Compute Cost ($/signing)              | 0.282 | 0.633  | 0.326       | 0.65   | 0.346         | 0.684  |
| Data Transfer (GiB/signing)           | 409.72| 819.44 | 409.72      | 819.44 | 409.72        | 819.44 |
| Data Transfer Cost ($/GiB)            | 0     | 0      | 0           | 0      | 0.02          | 0.02   |
| Total Data Transfer Cost ($/signing)  | 0     | 0      | 0           | 0      | 8.19          | 16.39  |
| Total Cost ($/signing)                | 0.282 | 0.633  | 0.326       | 0.65   | 8.54          | 17.07  |

**Table 1: Cost results.** The results shown are for generating a signature with the ECDSA on the curve secp192k1 using Yao's semi-honest garbled circuits protocol and Dual Execution. For the Local execution, there is a single host executing both sides of the protocol. Same Region uses two host in in the same EC2 region (US East); Long Distance uses one host in US East and the other in US West. Costs are rough estimates computed based on current AWS pricing, $0.398 per hour for an c4.2xlarge Amazon EC2 node. Bandwidth charges can vary by region, but are approximately $0.02/GiB at bulk transfer rates across regions within AWS.

These costs seem high, and what AWS charges is probably an over-estimate of what it would actually cost to do this at scale and bandwidth costs vary substantially across cloud providers. Nevertheless, we believe they are already low enough for joint signing to be practical for high-value certificates. For instance, Symantec, which charges a $500/year premium for ECC over RSA/DSA certificate, could easily absorb such bandwidth costs in the margins available for premium certificates, and compared to the human costs of validating identities for EV certificates these costs are minor.

**Cost Breakdown.** The garbled circuit used for signing contains 21.97 billion gates. Since we use the half-gates technique as implemented by Obliv-C, we need to send two 80-bit ciphertexts for each gate (160 bits total). This amounts to 409.18 GiB of total data transfer. The measured total bandwidth required for a signing is 409.72 GiB, indicating that all the other aspects of the protocol (including the OT to obtain input wires, output reveal, and other network overhead) are less than 0.54 GiB (<%0.132) of the total. The data transfer overhead is approximately double for dual execution protocol as it requires two rounds of Yao's protocol execution.

Since the garbled circuit transfer dominates the cost of decentralized signing, we break down the cost of its component operations in terms of number of non-XOR gates and the bandwidth to transfer them in Table 2. The multiplication and division operations over the 192-bit parameters are much more expensive than addition and subtraction. Then, multiplication and division operations make the modular inverse expensive as it relies on these primitives. The curve multiplication becomes expensive as it repetitively invokes the other operations, and the 19.2 Billion gates needed for this account for nearly the entire cost of the protocol, hence are the best target for optimization efforts.

**Signing across Cloud Services.** In most important scenarios, the certificate authorities would not want to rely on a single cloud service provider due to either trust or privacy issues or due to having legal obligations to a specific service provider. We performed experiments for signing between an AWS instance and an Azure instance in the same region (US East), both with same configuration. The cost to transfer data out of AWS service for the first 10 TB per month is 9 cents per GiB [59], and for Azure it is 8.7 cents per GiB [7]. The bandwidth between Azure and AWS US East nodes is 1.29 Gbits/sec, and the local runtime for an Azure node is similar to that of the Amazon node. We found that the Yao's protocol obtained the same optimal settings as that of two AWS nodes in the same region. For 32 simultaneous signings, Yao's protocol took 14.1 hours in comparison to 13.1 hours for two AWS nodes in same region. Thus, signing between Amazon and Azure nodes cost around $37. Though we did not perform signing experiments with dual execution due to cost constraints, we can estimate that running a signing with dual execution across Amazon and Azure nodes would cost around $74. Although the cost is higher than signing within the same cloud platform, this scenario is still practically feasible for professional certificate authorities that prioritize privacy over cost and can probably obtain bulk bandwidth during low-demand times at much cheaper rates than the costs we modeled.

| Operation            | # of Gates  | Bandwidth (in MiB) |
|----------------------|-------------|--------------------|
| Addition             | 1500        | 0.03               |
| Subtraction          | 1500        | 0.03               |
| Multiplication       | 101 000     | 1.93               |
| Division             | 376 000     | 7.17               |
| Modular Inverse      | 113 million | 2155.30            |
| Curve Multiplication | 19.2 billion| 366 210.94         |

**Table 2: Complexity of individual operations over 192-bit parameters in Garbled Circuit protocol**

## 6 DISCUSSION

We have demonstrated our decentralized certificate authority using secure two-party computation with the signing algorithm ECDSA on the curve secp192k1 as a proof-of-concept. The prototype can be expanded to work on positive curve parameters of any size without any change to the code, and it would be straightforward to implement other signing algorithms such as DSA this way (although the actual practicality of implementing other algorithms within an MPC would depend on the computation required). Below we mention some suggestions and possible modifications for efficiency and wider adaptability of our current scheme.

**Other Security Models.** Though we achieve secure two-party signing in the presence of malicious adversaries, our dual execution based signing costs double than that of Yao's protocol based signing. An alternative would be to use the more efficient protocol of Kolesnikov and Malozemoff [34], which provides a slightly weaker notion of publicly-verifiable covert security, but is practical and achieves efficiency close to that of semi-honest protocols. Covert adversaries [6] can deviate from the protocol like malicious adversaries, but are caught by the honest party with a fixed probability. Asharov and Orlandi [5] modified this definition by making the detection of covert adversaries publicly verifiable, thereby making this security notion more practical and widely applicable. This scheme is improved upon by Kolesnikov and Malozemoff, making it more efficient by replacing the original oblivious transfer [54] in the scheme with the more efficient oblivious transfer extension protocol [4]. The drawback of using this model, however, it provides no limit on the leakage that might be obtained before the adversary is caught. Without a strong revocation mechanism, this would have a high risk in our application since a malicious host could potentially obtain the other host's key share, and generate fraudulent certificates that would be trusted by browsers even if the malicious activity is detected.

**Extending to Three-Party Setting.** Our work has focused on the two-party setting, which we believe is simplest to deploy and reason about. There are very efficient MPC protocols know for three-party models where it is assumed that two of the parties are honest. For example, Araki et al. [3] propose a three-party scheme for distributed AES encryption and distributed ticket generation in Kerberos in the presence of a corrupt/malicious party. Launchbury et al. [37] and Laur et al. [38] also provide efficient implementations of distributed AES encryption with honest majority. Any of these schemes could be used for extending distributed certificate authority to three-party setting with an honest majority.

**Deployment.** In Section 3.4, we explained the setups for our decentralized CAs. Deployment of our designs would require some out-of-band changes on the current certificate chain and its validation process. First, browsers would need to add a new class of certificate, the joint validation certificate, analogous to the current extended validation certificate (EV) (Section 2.1).

In the first scenario of a single CA signing the certificate internally, the only change that the CA would need to make is to submit its new certificate to the browser. In the second scenario where two individual CAs are involved, the joint signing key would need to be trusted by the browser. Ideally, this could be a built-in root key in the browser so there is no reliance on a separate root key. For immediate deployment, however, it could be validated as part of a certificate chain.

In the third scenario, each subject-$CA_s$ pair has their own set of certificates. Since it would not be practical to include the certificate for every subject-$CA_s$ pair in the browser, we could use alternative measures to track subject-$CA_s$ certificates and ensure that CA-only certificates for the subject will not be trusted. The first idea is to use a certificate transparency log, where certificates are publicly logged by CAs and browsers as they are issued and observed as to easily audit and mitigate attempts of misuse [39]. A simpler solution is public key pinning [14], which associates a host directly with its certificate or public key before the deployment of the browser or when the browser first encounters the host. Once a subject has pinned its key using a subject-$CA_s$ certificate, it is protected with this client from a certificate generated independently by the CA.

## 7 RELATED WORK

We discuss existing works on distributing ECDSA signing, distributed certificate authority and multi-party AES computation which are closely related to our problem.

**Distributing ECDSA signing.** We describe two alternative protocols for distributed ECDSA signing. Threshold cryptography [19] allows $n$ parties to share the signing power so that any subset of $t + 1$ parties can jointly sign. Many schemes for producing standard digital signatures have been designed. Notably, Gennaro et al. [26] proposed a threshold-optimal, distributed DSA/ECDSA signature scheme that is fully compatible with the application of Bitcoin. This scheme requires constant computation time and storage for each party, taking around 13s to produce a signature.

Recently, Lindell [41] proposed a faster signature protocol for ECDSA and DSA for the specific case of signing with two parties. Their scheme is approximately two orders of magnitude faster than the previous best, taking 37 ms for signing with curve P-256 between two single core machines.

Though these schemes are much more efficient than ours, they are tailor-made to perform a specific signing protocol and hence lack the flexibility of performing complex multi-party assertions, such as the comparison of the hashed input messages of the two parties or checking a CA-specific policy, during the certificate signing process. Our scheme achieves this flexibility via the MPC protocol and hence could be easily extended to compute any custom signing.

**Distributing Certificate Authority.** Decentralizing or distributing a certificate authority across many nodes is an idea first introduced by Zhou and Haas [72] in 1999 when tackling the problem of providing security for ad hoc, mobile networks where hosts relied on each other to keep the network connected. This work proposed spreading the functionality of a single certificate authority to a set of nodes using secret sharing [61] and threshold cryptography. Later, researchers such as Dong et al. [21] built upon this idea and provided practical deployment solutions. Though we do not consider mobile ad hoc networks, we attempt to solve the same problem through secure multi-party computation [69].

**AES with Multi-Party Computation.** Multi-party computation has been used in many works [3, 18, 37, 38] for joint AES computation among multiple participants. Damgard and Keller [18] implemented three-party MPC protocol using VIFF framework for AES computation with honest majority for semi-honest adversary and achieved a block encryption in 2 seconds. The authors, however, claim that their method can be strengthened by using passively secure protocol [9] against dishonest majority or actively secure protocol [17] against a third of corrupted parties. Launchbury et al. [37] proposed a more efficient method for three-party AES encryption in the presence of semi-honest adversaries, where they replaced garbled circuits with look-up tables in the MPC protocol, and achieved 300 block encryptions per second. Laur et al. [38] implement multi-party AES with secret-shared key and plaintext using Sharemind framework [10], and then apply join operation over secret-shared tables in a database. Recently, Araki et al. [3] gave the first practical implementation of AES encryption with three-party protocol that provides security against semi-honest adversaries, and also guarantees privacy against malicious adversaries with honest majority. This privacy guarantee is weaker than security against malicious adversaries, in a sense, it only ensures that the malicious adversary does not infer about input or output of the computation, but can corrupt the computation, thereby affecting the correctness of the computation. Nonetheless, their method achieves 1 million AES operations per second. However, this work interests us the most since it also implements distributed ticket-granting ticket generation of Kerberos protocol under the same privacy settings, which is similar to our application of distributed certificate authority. Thus, extending the distributed certificate authority to three-party setting by adopting the method of Araki et al. is of particular interest, and is a motivation for future research.

## 8 CONCLUSION

Internet security depends crucially on certificates signed by trusted CA keys. These private keys pose a single point of failure, and it is essential to limit the risk that they are compromised or abused. We have demonstrated as a proof-of-concept that secure multi-party computation can be used to enable jointly-signed certificates without ever exposing the private signing key. Our experiments indicate that a decentralized CA is feasible and affordable today.

### Availability

Our code is available under an open source license from:

https://github.com/hainali/decentralizedca.

### ACKNOWLEDGMENTS

The authors thank Jack Doerner for providing the Absentminded Crypto Kit and extensive help with using it, and Samee Zahur for developing and supporting Obliv-C. The work was partially funded by an award from the National Science Foundation (1111781) and gifts from Amazon, Google, and Microsoft.

## A  ECDSA CODE IN OBLIV-C

Obliv-C framework is an extension of C and hence, the syntax of the code remains the same, but with some Obliv-C specific keywords and functions. The code snippet for ECDSA in Obliv-C is shown below.

```
1  void signCertificate(void * args) {
2      obig k_A, k_B, k, ...;
3      obliv uint8_t k_A_array[MAXN], k_B_array[MAXN], ...;
4      obig_init(&k_A, MAXN);
5      obig_init(&k_B, MAXN);
6      //... and so on
7
8      feedOblivCharArray(k_A_array, io->k_A, MAXN, 1);
9      feedOblivCharArray(k_B_array, io->k_B, MAXN, 2);
10     //... and so on
11
12     obig_import_opointed_be(&k_A, k_A_array, MAXN);
13     obig_import_opointed_be(&k_B, k_B_array, MAXN);
14     //... and so on
15
16     for (int i = 0; i < Z_LENGTH; i++) {
17         z_A[i] = ocBroadcastChar(io->z_A[i], 1);
18         z_B[i] = ocBroadcastChar(io->z_B[i], 2);
19     }
20
21     assert(compare(z_A, z_B, Z_LENGTH));
22     obig_import_pointed_be(&z, z_A, Z_LENGTH);
23
24     combine_shares(&sk, sk_A, sk_B);
25     combine_shares(&k, k_A, k_B);
26
27     curveMult(&R_x, &R_y, G_x, G_y, k, a, p);
28     obig_div_mod(&tmp, &r, R_x, n);
29     r_comp_result = obig_cmp(r, zero);
30
31     mult_inverse(&k_inverse, k, n);
32     obig_mul(&interim, r, sk);
33     obig_add(&interim, z, interim);
34     obig_div_mod(&tmp, &interim_mod, interim, n);
35     obig_mul(&interim, k_inverse, interim_mod);
36     obig_div_mod(&tmp, &s, interim, n);
37     s_comp_result = obig_cmp(s, zero);
38
39     revealOblivInt(&io->RisZero, r_comp_result, 0);
40     revealOblivInt(&io->SisZero, s_comp_result, 0);
41
42     for (int i = 0; i < r.digits; i++)
43         revealOblivChar(&io->r[i], r.data[i], 0);
44
45     for (int i = 0; i < s.digits; i++)
46         revealOblivChar(&io->s[i], s.data[i], 0);
47
48     obig_free(&k_A);
49     obig_free(&k_B);
50     //... and so on
51 }
```

Line 2 declares the 'obig' variables for storing curve parameters, signing keys and message to be signed. 'obig' variables can store any arbitrary length big numbers and support basic arithmetic operations like addition, subtraction, multiplication and division. We require 'obig' variables to store the 192-bit curve parameters for our implementation of 'secp192k1'. Line 3 declares 'obliv' variables which are used to temporarily store the random number and key shares. Lines 5–7 initialize the 'obig' variables. The content of these variables is not revealed to any party, unless revealObliv<Type>() function is invoked explicitly. Lines 9–15 load the random number and key shares from the two parties, such that $k_A$ is loaded from party 1 and $k_B$ is loaded from party 2, and so on.

Before the certificate signing process begins, the hash of 'tbsCert' from both the parties are to be negotiated. Both the parties load their hash values $z_A$ and $z_B$ in plain text (lines 17–19). Next, assertion is done in line 22, where we compare that both the hash values $z_A$ and $z_B$ are equal. The signing process continues only if the assertion passes, otherwise the protocol is terminated. Next, the private key shares and random number shares are combined in lines 25 and 26. ECDSA certificate signing is performed in lines 28–38 according to Section 2.2. After the computation, we reveal to both the parties if the $r$ or $s$ was zero (in lines 40 and 41), indicating the parties to rerun the algorithm with different shares of $k_A$ and $k_B$. Finally, signature pair $(r, s)$ is revealed to both the parties (lines 43–47), and this concludes the MPC protocol.

The same code can be easily switched between Yao's protocol and Dual Execution protocol. In order to execute Yao's protocol, 'signCertificate()' function is passed as an argument to 'execYaoProtocol()' function in a wrapper code. To run Dual Execution, 'signCertificate()' function is passed as an argument to 'execDualexProtocol()' function in the wrapper code.